\documentclass[10pt, conference, compsocconf]{IEEEtran}
\usepackage{multirow}
\usepackage{amsmath}
\usepackage{rotating}
\usepackage[table,usenames,dvipsnames]{xcolor}
\usepackage{graphicx,epsfig}
\usepackage{cite}
\usepackage{url}
\usepackage{color}
\usepackage{soul}
\usepackage{paralist}
\usepackage{booktabs}

\begin{document}

\title{Examining the Impact of Platform Properties on Quality Attributes}

\author{\IEEEauthorblockN{Balwinder Sodhi and T.V. Prabhakar}
\IEEEauthorblockA{Dept. of Computer Science and Engineering,
IIT Kanpur, UP 208016 India\\
\{sodhi, tvp\}@cse.iitk.ac.in}
}

\markboth{Journal of \LaTeX\ Class Files,~Vol.~6, No.~1, January~2012}%
{Shell \MakeLowercase{\textit{et al.}}: Bare Demo of IEEEtran.cls for Computer Society Journals}

\IEEEcompsoctitleabstractindextext{%

\begin{abstract}
We examine and bring out the architecturally significant characteristics of various virtualization and cloud oriented platforms. The impact of such characteristics on the ability of \textit{guest} applications to achieve various quality attributes (QA) has also been determined by examining existing body of architecture knowledge. We observe from our findings that efficiency, resource elasticity and security are among the most impacted QAs, and virtualization platforms exhibit the maximum impact on various QAs.
\end{abstract}

}

\maketitle

\IEEEdisplaynotcompsoctitleabstractindextext
\IEEEpeerreviewmaketitle

\section{Introduction}
\label{sec:intro}
Virtualization and cloud oriented computing platforms have emerged as quite disruptive candidates for deployment and development of software applications. For designing the architecture of a complex software system, it is critical to understand important properties of target computing platforms. Equally important is to know the impact of these properties on various functional and non-functional aspects of an application. Towards that end, we have examined and analysed architectural aspects of virtualization and cloud based platforms.

For the said platforms and their underlying components, we determine what are the architecturally significant \emph{characteristics} that are important from an application's perspective. We used the term \emph{characteristics} here to collectively represent both functional features and the non-functional quality present in the platform. The impact of such characteristics on different QAs of a guest application has been determined by examining software architecture body of knowledge and performing benchmarking experiments. This impact knowledge is critical for assessing different platforms on a set of QA criteria. 

\subsection{Related work}
Several researchers have explored different dimensions of virtualization and cloud based platforms. For instance, a dissection of the cloud into five main layers, and illustrating their interrelations and inter-dependency on constituent components has been discussed by Youseff et. al. \cite{youseff2008toward}. The architectural requirements of cloud platforms have been discussed by Rimal et. al. \cite{rimal2011architectural}. These works examine different technical dimensions of cloud computing. However, from the standpoint of an application that wants to make exploit capabilities of such modern computing platforms, some questions still do not have clear answers. For example:
\begin{enumerate}
\item What are important characteristics of various platforms from the viewpoint of a guest application?
\item How do such characteristics impact various functional and non-functional aspects of guest application?
\end{enumerate}

In the presented work we address the above questions. 

This report is organized into four sections. In Section \ref{sec:plat_chars} we examine the characteristics of various computing platforms. We bring out and discuss impact of said characteristics on QAs in Section \ref{sec:platform_impact}. Report is concluded in Section \ref{sec:concl}.

\section{Computing Platforms Characteristics}
\label{sec:plat_chars}
A platform has characteristics which are typically determined and specified by
\begin{itemize}
\item Its functional attributes.
\item Non-functional QAs that it assures.
\item Design decisions or tactics employed in its architecture.
\end{itemize}

The presented work examines in detail the characteristics of two platforms viz. \textit{Virtualization based} and \textit{Cloud based}. In order to bring out such characteristics, following documentation artifacts were examined:
\begin{itemize}
\item Architecture description documents. They provide different views, e.g. design decision view, deployment view etc., for platform's architecture.
\item Product specifications that describe functional and non-functional features.
\item Benchmarking data if available.
\end{itemize}
The platforms and their important properties have been discussed in detail in sections below. 
\subsection{Virtualization Based}
\label{sec:platform_virt}
In this type of platforms the hardware resources are virtual. Typically, the computing environment is offered as Virtual Machine (VM) which the Virtual Machine Monitor (VMM) executes on a physical hardware shared with other VMs.

Virtualization can be of two kinds: 
\begin{inparaenum}[a)]
\item OS based as in Solaris containers \cite{solaris_cont}, and 
\item Virtual Machine Monitor (VMM) based as in Xen \cite{xen_vmm}.
\end{inparaenum}
Further, the VMM based virtualization can be of two kinds: bare metal and hosted \cite{virt_overview_ibm}. Certain VMMs also exist that take advantage of special purpose hardware features such as of recent Intel processors \cite{intel_lvmm}. 

The internals of various types of virtualization platforms (such as \cite{vmm-sota-trends2005,virt_overview_ibm,libvirt2009,solaris_cont,xen_vmm,intel_lvmm}) were examined in detail to bring out their key properties. Following are the chief properties \emph{common} across above said types of virtualization platforms:
\begin{enumerate}
\item Limited hardware control capabilities for VMs
\item Programmatic self-serviced provisioning
\item VM check-pointing and snap shots
\item Software abstraction of hardware
\item Multi-tenancy by logical partitioning of physical host into encapsulated software entities
\item Abstraction of hardware platform specific APIs and ABIs (e.g. VMware's products export an x86-based computer)
\item VM migration (both live and offline)
\end{enumerate}

OS based virtualization platforms exhibit the following characteristics:
\begin{enumerate}
\item Guest OS (or kernel in some cases) in VM cannot be different from the host one.
\item Programs in VMs use the OS’s normal system call interface (no emulation involved).
\item Tight integration with host OS, and less overhead in comparison to the VMMs
\item Isolation under a shared OS instance only
\item Privileged tasks only under the host OS control (e.g. loading a device driver, changing IP in a VM); else the application has to be modified to work in a VM.
\item Uses file-level copy-on-write mechanisms
\end{enumerate}

Characteristics exhibited by VMM based virtualization platforms are as below:
\begin{enumerate}
\item No restrictions on guest OS in VM
\item Programs in VMs use VMM provided emulation instead of the OS’s normal system call interface into underlying hardware
\item Allows full guest OS control in the VM
\item Uses block-level copy-on-write
\item Bare-metal VMM:
	\begin{enumerate}
	\item In some cases modification of guest OS needed for running in a VM
	\item Thin and encapsulated hardware facing layer
	\end{enumerate}
\item Hosted VMM:
	\begin{enumerate}
	\item No modification needed to guest OS for running in a VM
	\item VM runs a regular process inside host OS subject to host OS environment
	\end{enumerate}
\end{enumerate}

\subsection{Cloud Based}
\label{sec:platform_cloud}
The functional features and available architecture and low-level implementation details of various cloud platforms were examined \cite{aws,euca2009,nimbus2009,one2011,xen_cloud,azure_learn,gae,chohan2010appscale} to bring out main characteristics of the said platforms. The key characteristics of cloud based computing platforms (common across various cloud variants) are listed as below:

\begin{enumerate}
\item Programmatic provisioning of resources
\item Allows self-service provisioning
\item Shared underlying computing infrastructure via multi-tenancy
\item Lack of standards for key services such as security, VM control and management among others.
\item Computing as a utility accessible over the network
\item Geographic location transparent to clients
\item Political/legal jurisdiction transparent to clients
\item Measured service
\item Lack of smart metering and billing. Users are billed on \emph{as used} basis.
\item Lack of absolute control on data and computing assets custody
\item Potential to abuse the relative anonymity behind registration and usage models
\item Lack of detailed and fine grained resource monitoring mechanisms
\item Difficult to assess software licensing structure, especially in complex deployment scenarios where multiple licenses of different software are used in a single environment.
\end{enumerate}

\medskip \textbf{Service Model Specific Characteristics:}

\subsubsection*{PaaS Cloud Specific}
\begin{enumerate}
\item Allows only provider supported programming languages, tools, APIs and components to build applications.
\item Can control deployed applications and possibly its hosting environment configurations.
\item No control of underlying infrastructure (network, servers, operating systems, or storage).
\end{enumerate}

\subsubsection*{IaaS Cloud Specific}
An IaaS cloud platform provides basic compute infrastructure as a VM plus some virtual storage and networking.
\begin{enumerate}
\item Allows resource utilization monitoring and reacting to events
\item Cloud user responsible for installing/managing all software on VM
\item Applications running in the VM are responsible for dealing with the reactions to above mentioned events. For instance, a configured reaction for a ``CPU utilization threshold reached event'' may be to add more instances of the VM. It is then expected that the architecture of the application(s) running on the VM allows to harness the newly added VM's capacity.
\item Limited control on networking components, e.g. host firewalls.
\end{enumerate}

\subsubsection*{SaaS Cloud Specific}
\begin{enumerate}
\item Allows control of a limited set of user-specific application configuration settings.
\item No control of underlying infrastructure (network, servers, operating systems, storage, or individual application capabilities).
\end{enumerate}

\medskip \textbf{Deployment Model Specific Characteristics:}

\subsubsection*{Public Cloud Specific}
\begin{enumerate}
\item Cloud service provider has the custody and control of applications, data and computing assets hosted on cloud.
\item Single point ownership of cloud infrastructure lies with the organization selling cloud services.
\item Often has homogeneous virtualization environment.
\item Allows limited configurations of cloud infrastructure.
\item Cloud infrastructure is made available to the general public for a fee.
\end{enumerate}

\subsubsection*{Private Cloud Specific}
\begin{enumerate}
\item Often has a homogeneous virtualization environment.
\item Total ownership, control and custody of applications, data and computing assets.
\item Allows custom configurations of cloud infrastructure.
\item Operated solely for one organization.
\end{enumerate}

\subsubsection*{Hybrid Cloud Specific}
\begin{enumerate}
\item A logical arrangement that combines two or more disparate clouds (private, community, or public).
\item Each constituent cloud remains a unique entity retaining its own characteristics.
\item Constituent clouds are linked via a technology (standardized or proprietary) that enables data and application portability.
\end{enumerate}

\subsubsection*{Community Cloud Specific}
\begin{enumerate}
\item Member organizations have total ownership, control and custody of applications, data and computing assets.
\item Supports a specific community that has shared goals/concerns.
\item Distributed ownership of cloud infrastructure shared by several participating organizations.
\item Often has a homogeneous virtualization environment.
\end{enumerate}

\section{Impact of Platform's Characteristics on QAs}
\label{sec:platform_impact}

The ability of any software system to achieve certain QAs is determined by:
\begin{enumerate}
\item Characteristics of the underlying platform on which the system is built
\item How the said characteristics have been harnessed (or mitigated) when designing architecture of the system.
\end{enumerate}
The characteristics that we refer to above are the result of design decisions taken by architects of the platform in question. Architecture of a well designed software system typically implements some architectural pattern(s). Architectural patterns employ proven design tactics for addressing design concerns and achieving the desired levels of QAs \cite{bass2003software,harrison2010architecture}.

Therefore, by examining platform characteristics in light of existing body of architecture knowledge, their impact on various QAs can be easily found. For our study of the QA impact of various platform characteristics, we chose a subset of QAs listed in  first part of the standard, ISO/IEC 9126-1. A partial list of QAs is:\\

\begin{tabular}{l l}
Assets custody	&	Operability	\\
Auditability	&	Performance isolation	\\
Availability	&	Policies Compliance	\\
Backup	&	Portability	\\
Configurability	&	Privacy	\\
Deployment	&	Reliability (MTBF)	\\
Disaster recovery	&	Resource elasticity	\\
Efficiency	&	Response time	\\
Environmental impact	&	Scalability	\\
Failure management	&	Security	\\
Installability	&	Supportability	\\
Interoperability	&	Tenent isolation	\\
Maintainability	&	Testability	\\
Modifiability	&	Throughput	\\
\end{tabular} \\

In subsequent sections, we have determined the impact of platform characteristics identified in Section \ref{sec:plat_chars} on above set of QAs. We examined in detail the architecture patterns, design tactics and best practices knowledge from sources such as \cite{gamma2002design,bass2003software,posa_vol2,posa_vol3,posa_vol5,alur2003core,abowd1997recommended} in order to determine the said impact on QAs. The information about impact of computing platform's characteristics on QAs is presented in the tabular (matrix) form. Along the rows are listed characteristics of the platforms, and along columns are indicated QAs. 

Impact of a platform characteristic on a QA can be located at the intersection of respective row and column. A value of 1 for impact indicates that the characteristic in selected row has favourable impact on the QA in the column. Value -1 indicates an adverse impact on the QA, and the empty cell means that there is no impact.

\begin{table*}
\scriptsize
\setlength{\tabcolsep}{2 pt}
\centering
\caption{QA impact for Virtualization platforms \label{tab:nfqa_impact_virt}}
\begin{tabular}{|p{24em}|*{28}{p{1 em}|}}
\rowcolor[gray]{.9}
\multicolumn{1}{p{15em}|}{\textbf{Characteristics $\Downarrow$}} & \rotatebox{90}{Auditability} & \rotatebox{90}{Availability} & \rotatebox{90}{Backup} & \rotatebox{90}{Configurability} & \rotatebox{90}{Deployment} & \rotatebox{90}{Disaster recovery} & \rotatebox{90}{Efficiency} & \rotatebox{90}{Environmental impact} & \rotatebox{90}{Failure management} & \rotatebox{90}{Installability} & \rotatebox{90}{Interoperability} & \rotatebox{90}{Maintainability} & \rotatebox{90}{Modifiability} & \rotatebox{90}{Operability} & \rotatebox{90}{Policies Compliance} & \rotatebox{90}{Portability} & \rotatebox{90}{Privacy} & \rotatebox{90}{Reliability (MTBF)} & \rotatebox{90}{Response time} & \rotatebox{90}{Scalability} & \rotatebox{90}{Security} & \rotatebox{90}{Supportability} & \rotatebox{90}{Testability} & \rotatebox{90}{Throughput} & \rotatebox{90}{Resource elasticity} & \rotatebox{90}{Assets custody} & \rotatebox{90}{Tenant isolation} & \rotatebox{90}{Performance isolation}\\ \hline 

\multicolumn{29}{l}{\textbf{Common across all platforms}} \\ \toprule
Software abstraction of hardware & 1 & 1 & 1 & 1 & 1 & 1 & 1 & 1 & 1 & 1 & 1 & 1 & 1 & 1 & 1 &  &  & 1 & -1 & 1 &  & 1 & 1 & -1 & 1 &  &  & -1 \\ \hline
Limited hardware control capabilities for Vms &  &  &  &  &  &  & -1 & -1 &  &  &  &  &  &  &  &  & -1 & -1 &  &  &  & -1 &  &  &  &  &  &  \\ \hline
Programmatic self-serviced provisioning &  & 1 & 1 &  & 1 & 1 &  & 1 & 1 & 1 &  &  &  &  &  &  &  & 1 &  & 1 &  & 1 &  &  & 1 &  &  &  \\ \hline
Multi-tenancy by logical partitioning of physical host &  &  &  &  & 1 & 1 & 1 & 1 & 1 & 1 & 1 &  &  &  &  & 1 & 1 & 1 & 1 & 1 & 1 & 1 & 1 & 1 &  &  & 1 & -1 \\ \hline
VM check-pointing and snap shots &  & 1 & 1 &  & 1 & 1 & 1 & 1 & 1 & 1 &  & 1 & 1 &  &  & 1 &  & 1 &  & 1 & 1 & 1 & 1 & 1 & 1 &  &  &  \\ \hline
VM migration (both live and offline) &  & 1 & 1 &  & 1 & 1 & 1 & 1 & 1 & 1 &  & 1 &  &  &  & 1 &  & 1 & 1 & 1 & 1 & 1 &  & 1 & 1 &  &  &  \\ \hline
Abstraction of platform specific APIs and ABIs &  &  &  &  &  &  &  &  &  & 1 & 1 & 1 & 1 &  &  & 1 &  &  & -1 &  &  &  &  & -1 & 1 &  &  &  \\ \hline

\multicolumn{29}{l}{\textbf{VMM Virtualization platform}} \\ \toprule
Allows full guest OS control in the VM & 1 & 1 &  & 1 & 1 & 1 &  &  &  &  &  &  &  &  & 1 &  &  &  &  &  &  &  &  &  & 1 &  &  &  \\ \hline
Uses block-level copy-on-write &  &  & -1 &  &  &  &  &  &  &  &  &  &  &  &  &  &  &  & -1 &  &  &  &  &  &  &  &  &  \\ \hline
No restrictions on guest OS in VM &  &  &  & 1 &  &  &  &  &  &  &  &  &  &  &  &  &  &  &  &  &  &  &  &  & 1 &  &  &  \\ \hline
Programs in VMs use VMM provided emulation instead of the OS’s normal system call interface &  &  &  &  &  &  & -1 &  &  &  &  &  &  &  &  &  &  &  & -1 &  &  &  &  & -1 &  &  &  & -1 \\ \hline

\multicolumn{29}{l}{\textbf{Bare-metal VMM platform}} \\ \toprule
Thin and encapsulated h/w facing layer &  &  &  &  &  &  & 1 &  &  &  &  &  &  &  &  &  &  &  & 1 & 1 &  &  &  & 1 &  &  &  &  \\ \hline

\multicolumn{29}{l}{\textbf{Hosted VMM platform}} \\ \toprule
VM runs a regular process inside host OS & 1 & -1 &  & 1 & 1 & 1 &  &  & 1 & 1 &  &  &  & 1 & 1 & 1 & -1 &  & -1 &  &  &  &  &  &  &  &  & -1 \\ \hline
Can run unmodification guest OS in a VM &  &  &  &  &  &  &  &  &  & 1 &  &  &  &  &  & 1 &  &  &  &  &  &  &  &  &  &  &  &  \\ \hline

\multicolumn{29}{l}{\textbf{OS Virtualization platform}} \\ \toprule
Isolation under a shared OS instance only &  &  &  & -1 &  &  & 1 &  & -1 &  &  &  &  & 1 & 1 &  & -1 & -1 & 1 &  & -1 & 1 & 1 & 1 &  &  &  &  \\ \hline
Privileged tasks only under the host OS control & -1 &  &  & -1 & -1 &  &  &  & -1 & -1 &  &  &  &  &  &  &  &  &  &  & 1 & -1 &  &  &  &  &  &  \\ \hline
Programs in VMs use the OS’s normal system call interface &  &  &  &  &  &  &  &  &  &  &  &  & -1 &  &  &  &  &  & 1 &  &  &  &  & 1 &  &  &  &  \\ \hline
Tight integration with host OS &  &  &  &  &  &  &  &  &  &  &  & -1 & -1 &  &  & -1 &  &  &  &  &  &  &  &  &  &  &  &  \\ \hline
Guest OS (or kernel in some cases) in VM must be same as host. &  &  &  & -1 &  &  &  &  &  &  &  &  & -1 &  &  & -1 &  &  &  &  &  &  &  &  &  &  &  &  \\ \hline
Uses file-level copy-on-write mechanisms &  &  &  &  &  &  &  &  &  &  &  &  &  &  &  &  &  &  &  & 1 &  &  &  &  &  &  &  &  \\ \hline

\end{tabular}
\end{table*}

\begin{table*}
\scriptsize
\setlength{\tabcolsep}{2 pt}
\centering
\caption{QA impact of cloud platforms \label{tab:nfqa_impact_cloud}}
\begin{tabular}{|p{24em}|*{28}{p{1 em}|}}
\rowcolor[gray]{.9}
\multicolumn{1}{p{15em}|}{\textbf{Characteristics $\Downarrow$}} & \rotatebox{90}{Auditability} & \rotatebox{90}{Availability} & \rotatebox{90}{Backup} & \rotatebox{90}{Configurability} & \rotatebox{90}{Deployment} & \rotatebox{90}{Disaster recovery} & \rotatebox{90}{Efficiency} & \rotatebox{90}{Environmental impact} & \rotatebox{90}{Failure management} & \rotatebox{90}{Installability} & \rotatebox{90}{Interoperability} & \rotatebox{90}{Maintainability} & \rotatebox{90}{Modifiability} & \rotatebox{90}{Operability} & \rotatebox{90}{Policies Compliance} & \rotatebox{90}{Portability} & \rotatebox{90}{Privacy} & \rotatebox{90}{Reliability (MTBF)} & \rotatebox{90}{Response time} & \rotatebox{90}{Scalability} & \rotatebox{90}{Security} & \rotatebox{90}{Supportability} & \rotatebox{90}{Testability} & \rotatebox{90}{Throughput} & \rotatebox{90}{Resource elasticity} & \rotatebox{90}{Assets custody} & \rotatebox{90}{Tenant isolation} & \rotatebox{90}{Performance isolation}\\ \hline 

\multicolumn{29}{l}{\textbf{Common across all cloud platforms}} \\ \toprule
Relative anonymity behind subscription and usage &  &  &  &  &  &  &  &  &  &  &  &  &  &  &  &  & -1 &  &  &  & -1 &  &  &  &  &  &  &  \\ \hline
Programmatic provisioning of resources &  & 1 & 1 & 1 & 1 & 1 & 1 & 1 & 1 & 1 &  &  &  & 1 &  &  &  & 1 & 1 & 1 &  & 1 & 1 & 1 & 1 &  &  &  \\ \hline
Allows self-service provisioning &  & 1 & 1 & 1 & 1 & 1 & 1 & 1 & 1 & 1 &  &  &  & 1 &  &  &  & 1 & 1 & 1 & -1 & 1 & 1 & 1 & 1 &  &  &  \\ \hline
Lack of absolute control on software/data assets custody & -1 & -1 &  &  &  &  &  &  & -1 &  &  &  &  &  & -1 &  & -1 &  &  &  & -1 &  &  &  &  &  &  &  \\ \hline
Lack of standards for key services & -1 &  &  &  &  &  &  &  &  &  & -1 & -1 & -1 &  & -1 & -1 &  &  &  &  &  &  &  &  &  &  &  &  \\ \hline
Measured service &  &  &  &  &  &  &  &  &  &  &  &  &  &  &  &  &  &  &  &  &  &  &  &  &  &  &  &  \\ \hline
Lack of smart metering and billing. Users billed on as used basis &  &  &  &  &  &  &  &  &  &  &  &  &  &  &  &  &  &  &  &  &  &  &  &  & -1 &  &  &  \\ \hline
Lack of fine grained resource monitoring mechanisms & -1 &  &  &  &  & -1 &  &  & -1 &  &  & -1 &  &  & -1 &  &  & -1 &  &  & -1 & -1 &  &  &  &  &  &  \\ \hline
Computing as a utility accessible over the network &  &  &  &  & 1 & 1 & 1 & 1 &  &  &  &  &  &  &  &  &  &  &  &  &  &  &  &  & 1 &  &  &  \\ \hline
Geographic location transparent to clients &  &  &  &  &  &  &  &  &  &  &  &  &  &  & -1 &  & -1 &  &  &  & -1 &  &  &  &  &  &  &  \\ \hline
Political/legal jurisdiction transparent to clients &  &  &  &  &  &  &  &  &  &  &  &  &  &  & -1 &  & -1 &  &  &  & -1 &  &  &  &  &  &  &  \\ \hline
Difficult to assess software licensing structure &  &  &  &  &  &  &  &  &  &  &  &  &  &  &  &  &  &  &  &  &  &  &  &  &  &  &  &  \\ \hline

\multicolumn{29}{l}{\textbf{PaaS platform}} \\ \toprule
Only deployed applications and its hosting environment configurations can be controlled & -1 &  &  & -1 &  &  &  &  & -1 &  &  &  &  &  &  &  &  &  &  &  &  &  &  &  &  &  &  &  \\ \hline
Only provider supported programming languages, tools, APIs etc. Allowed & -1 &  &  &  &  &  &  &  &  &  & -1 & 1 & -1 &  &  & -1 &  &  &  &  &  &  &  &  &  &  &  &  \\ \hline
No control/responsibility of underlying infrastructure (hardware, OS) & -1 &  &  & -1 &  &  &  &  &  & 1 &  &  &  &  &  &  & -1 &  &  &  & -1 &  &  &  &  &  &  &  \\ \hline

\multicolumn{29}{l}{\textbf{SaaS platform}} \\ \toprule
Can control only a limited set of user-specific application configuration settings. & -1 &  &  & -1 &  &  &  &  &  &  &  &  & -1 &  &  &  &  &  &  &  &  &  &  &  &  &  &  &  \\ \hline
Provider specific service implementation &  &  &  &  &  &  &  &  &  &  &  &  &  &  &  & -1 & -1 &  &  &  &  &  &  &  &  &  &  &  \\ \hline

\multicolumn{29}{l}{\textbf{IaaS platform}} \\ \toprule
User responsible for installing/managing all software on VM &  &  &  &  &  &  &  &  &  & 1 &  &  &  & -1 &  &  &  &  &  &  &  &  &  &  &  &  &  &  \\ \hline
Allows resource utilization monitoring and reacting to events &  & 1 &  & 1 &  &  & 1 & 1 & 1 &  &  &  &  &  & 1 &  &  & 1 &  & 1 & 1 &  &  & 1 & 1 &  &  &  \\ \hline
Applications in VM are responsible for handling above events &  &  &  &  &  &  &  &  & 1 &  &  &  &  &  &  & -1 &  &  &  & 1 &  &  &  &  &  &  &  &  \\ \hline
Limited control on networking components  &  &  &  & -1 &  &  &  &  &  &  &  &  &  &  &  &  &  &  &  &  &  &  &  &  & -1 &  &  &  \\ \hline


\multicolumn{29}{l}{\textbf{Public Cloud platform}} \\ \toprule
Infrastructure available to general public for fee &  &  &  &  &  &  &  &  &  &  &  &  &  &  &  &  & -1 &  &  &  & -1 &  &  &  &  &  &  &  \\ \hline
Cloud vendor has single point ownership of infrastructure &  &  &  &  &  &  &  &  &  &  &  &  &  &  &  &  &  &  &  &  & -1 &  &  &  &  &  &  &  \\ \hline
Cloud provider retains custody and control of software/data assets &  &  &  &  &  &  &  &  &  &  &  &  &  &  &  &  & -1 &  &  &  &  &  &  &  &  & -1 &  &  \\ \hline

\multicolumn{29}{l}{\textbf{Private Cloud platform}} \\ \toprule
Operated solely for one organization &  &  &  &  &  &  &  &  &  &  &  &  &  &  &  &  &  &  &  &  &  &  &  &  &  & 1 &  &  \\ \hline
Total ownership/custody of software/data assets & 1 &  &  &  &  & 1 &  &  &  &  &  &  &  &  &  &  & 1 &  &  &  & 1 &  &  &  &  & 1 &  &  \\ \hline
Custom configurations of cloud infrastructure possible  & 1 & 1 &  &  &  &  & 1 &  &  &  &  &  &  &  &  &  &  &  &  &  &  &  &  &  &  &  &  &  \\ \hline
Has a homogeneous virtualization environment &  &  &  &  &  & 1 & 1 &  &  &  & 1 &  &  &  & 1 &  &  &  &  &  &  &  &  &  & 1 &  &  &  \\ \hline

\multicolumn{29}{l}{\textbf{Hybrid Cloud platform}} \\ \toprule
Blending of two or more disparate clouds &  &  &  &  &  &  &  &  &  &  &  &  &  &  &  &  & -1 &  &  &  & -1 &  &  &  &  & -1 &  &  \\ \hline

\multicolumn{29}{l}{\textbf{Community Cloud platform}} \\ \toprule
Ownership of cloud infrastructure shared by participating organizations &  &  &  &  &  &  &  &  &  &  &  &  &  & -1 &  &  &  &  &  &  &  &  &  &  &  &  &  &  \\ \hline
Has a homogeneous virtualization environment &  &  &  &  &  & 1 & 1 &  &  &  & 1 &  &  &  & 1 &  &  &  &  &  &  &  &  &  & 1 &  &  &  \\ \hline
Member organizations have ownership and custody of software/data assets  &  &  &  &  &  &  &  &  &  &  &  &  &  &  &  &  & 1 &  &  &  & 1 &  &  &  &  & 1 &  &  \\ \hline

\end{tabular}
\end{table*}

\subsection{Impact of Virtualization Based Platform Characteristics}
\label{sec:impact_virt}
Impact information for only a subset of all possible combinations between platform characteristics and QAs are described in detail here. The Table \ref{tab:nfqa_impact_virt}, however, shows the impact information for all cases. It is easy to observe that as we move from parent to the subcategories of virtualization types such as VMM based, OS based and further down, the impact on QAs gets narrowed down to fewer QAs due to specialization.

\subsubsection{Ability to Take VM Snapshots}
Virtualization provides several capabilities to enable programmatic manipulation of VMs. It allows saving the state of a live or off-line VM to a file by taking a snapshot of the VM. This is similar to check-pointing of in-fight transactions in database systems. One can easily restore the VM to a prior good known state by restoring a snapshot on detecting a failure. This favourably impacts reliability, disaster recovery, backup and deployment etc. Table \ref{tab:nfqa_impact_virt} shows impact on rest of the QAs.

\subsubsection{Abstraction of Hardware Resources as Software Entities}
The hardware resources such as CPU, memory, disk etc., are software entities in a virtualization based platform. For instance, CPU seen inside the VM is a representation of time-slices on underlying physical CPU cores. Design tactic of \emph{abstraction} has been applied to present physical resources as software entities to the applications. This idea allows the VMs to be programmatically examined, controlled, saved and moved around over the network like regular files. Almost all the QAs, except some performance related QAs, that we listed are favourably impacted by this characteristic. For instance, backup, disaster recovery, operability etc. are favourably impacted. On the other hand, the additional layer of abstraction introduces performance penalties; as such response time and throughput are expected to be adversely impacted. 

\subsection{Impact of Cloud Based Platform Characteristics}
Several variants of cloud based platforms exist, mainly differing based on service (i.e. IaaS, PaaS etc.) and deployment models (i.e. private, public etc.). They have many of the characteristics common, whereas some are specific to each variant. There are large number of combinations possible between platform characteristics and QAs. To highlight our analysis approach, we discuss the impact of only a subset of characteristics on QAs. Impact information for rest of the combinations is, however, presented in Table \ref{tab:nfqa_impact_cloud}.

\subsubsection{Limited Control of Underlying Platform}
The cloud users get only a limited control on underlying platform infrastructure in all service model based non-private variants of cloud. For instance, IaaS host machine's power management functions are typically not available in VMs. That is, a VM is not allowed to put the physical CPU to a low power state when idle. Similarly, on PaaS, only deployed applications and its hosting environment configurations can be controlled by applications. As such, this characteristic adversely impacts adaptability, configurability and failure management. Most of the remaining QAs are not directly impacted and are as shown in Table \ref{tab:nfqa_impact_cloud}.

\subsubsection{Self-service Provisioning}
Cloud platforms provide programmatic APIs to allow automation of several common provisioning tasks. For instance, a VM with 4 CPUs, 8GB RAM, 250GB of disk can be created in seconds. This characteristic improves deployment, configurability, operability, backup etc. This is because achieving these QAs now does not require human/manual intervention. These tasks can now be performed via programmatic means by utilizing cloud APIs. We have determined the impact on other QAs by a similar analysis, and is shown in Table \ref{tab:nfqa_impact_cloud}.

\section{Conclusion}
\label{sec:concl}
We have presented the detailed examination and analysis of various platforms, from the standpoint of guest application's architecture and design. Both virtualization and cloud platforms possess characteristics that impact the ability of guest applications to achieve certain QAs. For instance, the ability to take a snapshot of a running VM impacts the disaster recovery in a favourable manner. Similarly, the lack of physical custody of data and software assets by the users in case of cloud platforms impacts the security and privacy QAs adversely. We also observe that certain QAs such as resource elasticity, tenant isolation and performance isolation arise mainly in case of cloud platforms.

We believe that these knowledge artifacts presented here will help in performing the rational technical evaluation of various platforms.


\end{document}